\documentclass[prl,aps,amssymb,showpacs,twocolumn]{revtex4}
\usepackage{graphicx}% Include figure files

\begin{document}

\title{Spin Charge Separation in the Quantum Spin Hall State}

\author{Xiao-Liang Qi and Shou-Cheng Zhang}

 \affiliation{Department of Physics, McCullough
Building, Stanford University, Stanford, CA 94305-4045}

\date{\today}

\begin{abstract}
The quantum spin Hall state is a topologically non-trivial insulator
state protected by the time reversal symmetry. We show that such a
state always leads to spin-charge separation in the presence of a
$\pi$ flux. Our result is generally valid for any interacting
system. We present a proposal to experimentally observe the
phenomenon of spin-charge separation in the recently discovered
quantum spin Hall system.

\end{abstract}


\maketitle

Spin-charge separation is one of the deepest concepts in condensed
matter physics. In the Su-Schrieffer-Heeger model of
polyacetylene\cite{su1979}, a domain wall induces two mid-gap
states, one for each spin orientation of the electron. If both
states are unoccupied, or both states are occupied, the domain
wall soliton has charge $\pm e$ but no spin. If only one of the
state is occupied, the domain wall soliton has spin $S_z=\pm 1/2$
but no charge. In this remarkable way, the two fundamental degrees
of freedom of an electron is split apart. Since then, the concept
of spin-charge separation has become a corner stone in condensed
matter physics. This phenomenon occurs generally in interacting
quantum many-body systems in one dimension, and can be
demonstrated convincingly by the bosonization techniques. The
concept has also been generalized to two dimensions. In
particular, it is conjectured that such a phenomenon occurs in the
high temperature superconductors\cite{kivelson1987,anderson1987}.
However, this phenomenon has not yet been convincingly observed in
any two dimensional systems.

Recently, a new two dimensional quantum state of matter has been
theoretically proposed\cite{kane2005a,bernevig2006b,bernevig2006}.
The quantum spin Hall (QSH) state is a topologically non-trivial
state of matter protected by the time reversal symmetry. It has a
bulk insulating gap, but has helical edge states on the sample
boundary, where electron states with opposite spins
counter-propagate at a given edge. This novel quantum state of
matter has recently been theoretically
predicted\cite{bernevig2006} and experimentally
observed\cite{koenig2007} in the $HgTe$ quantum wells. The
topological property of the quantum Hall (QH) state is described
by an integer Chern number\cite{thouless1982}, defined over the
single particle momentum space, and this integer is directly
related to the experimentally observed quantum of Hall
conductance. This construction can also be generalized to an
interacting system, where the Chern number is defined over the
space of twisted boundary conditions\cite{niu1985}. The
topological property of the QSH state is currently described by a
$Z_2$ topological
number\cite{kane2005b,moore2007a,roy2006b,sslee2007}, which is
also defined over the single particle momentum space. This $Z_2$
classification has provided an important insight on the
topological non-triviality of the QSH state. However, unlike the
situation in QH systems, there are several fundamental missing
links in the QSH systems. Needed is a general classification of
time reversal invariant (TRI) topological insulators in two
dimensions which is valid in the presence of arbitrary
interactions. Such a general classification beyond the single
particle band picture is especially called for since the concept
of a topological Mott insulator has recently been
introduced\cite{raghu2007}. More importantly, we need to find
experimentally measurable properties which directly demonstrate
the topological non-triviality of the QSH state.

In this paper, we solve both problems by providing a deep connection
between the concept of spin-charge separation and the QSH effect.
Following Laughlin's argument for the QH effect, we consider the
adiabatic insertion of a pure gauge flux in the QSH state. We show
that there are four different ways of reaching the final flux of
$\pi$, and that these four processes create the spin-charge
separated holon, chargeon and two spinon states which are
exponentially localized near the flux. We then prove two general
theorems providing a $Z_2$ classification of TRI insulators in two
dimensions. This new classification scheme is generally valid in the
presence of many-body interactions, and leads to spin-charge
separation as its direct physical consequence. Finally, we propose
an experimental setting to observe the phenomenon of spin-charge
separation in the recently discovered QSH system.

\begin{figure}[htpb]
    \begin{center}
        \includegraphics[width=3in]{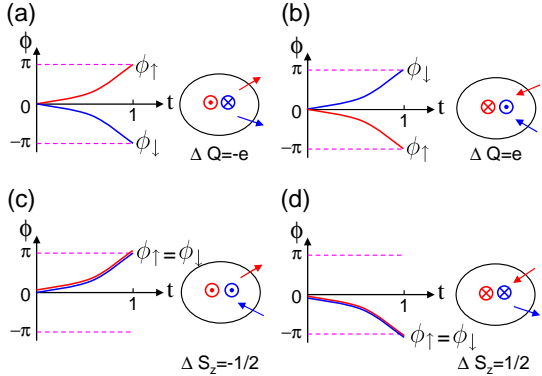}
    \end{center}
    \caption{Four different adiabatic processes from
    $\phi_\uparrow=\phi_\downarrow=0$ to $\phi_\uparrow=\phi_\downarrow=\pm\pi$. The red (blue) curve stands for
    the flux $\phi_{\uparrow(\downarrow)}(t)$, respectively. The symbol ``$\odot$" (``$\otimes$") represents
    increasing (decreasing) fluxes, and the arrows show the current
    into and out of the Gaussian loop, induced
    by the changing flux. Charge is pumped in the processes with
    $\phi_\uparrow(t)=-\phi_\downarrow(t)$, while spin is pumped in those with
    $\phi_\uparrow(t)=\phi_\downarrow(t)$.}
    \label{fig:spinonholon}
\end{figure}

We first present an argument which is physically intuitive, but
only valid when there is at least a $U_s(1)$ spin rotation
symmetry. In this case, the QSH effect is simply defined as two
copies of QH, with opposite Hall conductances of $\pm e^2/h$ for
opposite spin orientations. Without loss of generality, we first
consider a disk geometry with a gauge flux of
$\phi_\uparrow=\phi_\downarrow=hc/2e$, or simply $\pi$ in units of
$\hbar=c=e=1$, through a hole at the center, see Fig.
\ref{fig:spinonholon}. The gauge flux acts on both spin
orientations, and the $\pi$ flux preserves time reversal symmetry.
We consider adiabatic processes of $\phi_\uparrow(t)$ and
$\phi_\downarrow(t)$, where
$\phi_\uparrow(t)=\phi_\downarrow(t)=0$ at $t=0$, and
$\phi_\uparrow(t)=\phi_\downarrow(t)=\pm\pi$ at $t=1$. Since the
flux of $\pi$ is equivalent to the flux of $-\pi$, there are four
different adiabatic processes all reaching the same final flux
configuration. In process (a),
$\phi_\uparrow(t)=-\phi_\downarrow(t)$ and
$\phi_\uparrow(t=1)=\pi$. In process (b),
$\phi_\uparrow(t)=-\phi_\downarrow(t)$ and
$\phi_\uparrow(t=1)=-\pi$. In process (c),
$\phi_\uparrow(t)=\phi_\downarrow(t)$ and
$\phi_\uparrow(t=1)=\pi$. In process (d),
$\phi_\uparrow(t)=\phi_\downarrow(t)$ and
$\phi_\uparrow(t=1)=-\pi$. These four processes are illustrated in
Fig \ref{fig:spinonholon}. Note that process (a) and (b) preserves
time reversal symmetry at all intermediate stages, while process
(c) and (d) only preserves the time reversal symmetry at the final
stage.

%\begin{figure}[htpb]
%    \begin{center}
%        \includegraphics[width=2in]{flux.eps}
%    \end{center}
%    \caption{Schematic picture of flux threading into a QSH system with conserved $S_z$. $\phi_\uparrow$ and $\phi_\downarrow$
%    are the flux for spin up and spin down electrons, and ${\bf j}_{\uparrow},{\bf j}_{\downarrow}$ are the current induced when
%    flux is changed adiabatically. The black dash line shows a Gaussian surface with radius $r_G$.}
%    \label{fig:flux}
%\end{figure}

We consider a Gaussian loop surrounding the flux. As the flux
$\phi_\uparrow(t)$ is turned on adiabatically, Faraday's law
induction states that a tangential electric field ${\bf
E}_\uparrow$ is induced along the Gaussian loop. The quantized
Hall conductance implies a radial current ${\bf
j}_\uparrow=\frac{e^2}{h} {\bf z}\times {\bf E}_\uparrow$,
resulting in a net charge flow $\Delta Q_\uparrow$ through the
Gaussian loop:
\begin{eqnarray}
\Delta Q_\uparrow & = & - \int_0^1 dt \int d {\bf n}\cdot {\bf
j}_\uparrow = - \frac{e^2}{h} \int_0^1 dt \int d {\bf l}\cdot {\bf
E}_\uparrow \nonumber\\
& = & - \frac{e^2}{hc} \int_0^1 dt  \frac{\partial \phi}{\partial
t} = - \frac{e^2}{hc} \frac{hc}{2e}=-\frac{e}{2}
\end{eqnarray}
Identical argument applied to the down spin component shows that
$\Delta Q_\downarrow=-e/2$. Therefore, this adiabatic process
creates the holon state with $\Delta Q=\Delta Q_\uparrow+\Delta
Q_\downarrow=-e$ and $\Delta S_z=\Delta Q_\uparrow-\Delta
Q_\downarrow=0$.

Applying similar arguments to process (b) gives $\Delta
Q_\uparrow=\Delta Q_\downarrow=e/2$, which leads to a chargeon
state with $\Delta Q=e$ and $\Delta S_z=0$. Process (c) and (d)
give $\Delta Q_\uparrow=-\Delta Q_\downarrow=e/2$ and $\Delta
Q_\uparrow=-\Delta Q_\downarrow=-e/2$ respectively, we therefore
obtain the spinon states with $\Delta Q=0$ and $\Delta S_z=\pm
1/2$. The Hamiltonians $H(t)$ in the presence of the gauge flux
are the same at $t=0$ and $t=1$, but differ in the intermediate
stages of the four adiabatic processes. Assuming that the ground
state is unique at $t=0$, we obtain four final states at $t=1$,
which are the holon, chargeon and the two spinon states. Both the
spin and the charge quantum numbers are sharply defined quantum
numbers\cite{kivelson1982}. The insulating state has a bulk gap
$\Delta$, and an associated coherence length $\xi=\hbar
v_F/\Delta$. As long as the radius of the Gaussian loop $r_G$ far
exceeds the coherence length, {\it i.e.}, $r_G \gg \xi$, the spin
and the charge quantum numbers are sharply defined within
exponential accuracy. Recently, similar proposals of
fractionalization phenomena in two-dimensions induced by
topological defects have been studied in several other
systems.\cite{lee2007,hou2007,seradjeh2007b,weeks2007}

While the argument above is intuitive and generally valid in the
presence of both interaction and disorder, it has a serious
shortcoming. It relies on the $U_s(1)$ spin rotation symmetry
which is not generic in the presence of spin-orbit interactions.
Therefore, we first need a general definition of the concept of
spin-charge separation relying only on the generic time-reversal
symmetry. In the absence of spin rotational symmetry, we can still
use the generic time reversal symmetry and the Kramers theorem to
classify integer versus half-integer spin states. Denoting the
time reversal operator as $\cal T$, and the charge operator as
$N$, we give the following general definition of spin-charge
separation:

\begin{itemize}\item\textsc{Definition I}: A generalized chargeon (or holon) state is
a quantum state $|\psi_c\rangle$ satisfying $(-1)^N |\psi_c\rangle
= - |\psi_c\rangle$, and ${\cal T} |\psi_c\rangle =
|\psi_c\rangle$. A generalized spinon state is a doublet of
quantum states $|\psi_s^+\rangle$ and $|\psi_s^-\rangle$,
satisfying $(-1)^N |\psi_s^\pm\rangle = |\psi_s^\pm\rangle$,
${\cal T} |\psi_s^+\rangle = |\psi_s^-\rangle$ and ${\cal T}
|\psi_s^-\rangle = -|\psi_s^+\rangle$.\end{itemize}

The Kramers degeneracy is generally lifted in the presence of a
magnetic field, and the resulting energy splitting of the doublet is
linear in magnetic field, {\i.e.} $\Delta E=g^* \mu_B |B|$. The
constant of proportionality $g^*$ can be defined as the effective
$g$ factor of the spinon. We now consider a TRI insulator without
any additional spin rotational symmetry. We consider the
generalizations of processes (a) and (b), by replacing the hopping
matrix elements $t_{ij}$ with $t_{ij}e^{i\theta(t)
\Gamma}$\cite{sheng2006,qi2006b} on all links along a string
extending from the flux tube to infinity. Here $\Gamma$ is a matrix
in the spin space, and the intuitive discussion given above
corresponds to the choice of $\Gamma=S_z$. All the following
discussions are valid even if $\Gamma$ is not conserved. We focus on
the case with $\Gamma$ an odd operator under the time reversal
symmetry, {\it i.e.} ${\cal T}^{-1} \Gamma {\cal T} = -\Gamma$.

\begin{itemize}\item\textsc{Theorem I}: For any time-reversal odd $\Gamma$ satisfying
$e^{i\pi \Gamma}=-1$, an integer number of charge $N_\Gamma e$ is
pumped towards the isolated flux tube during the adiabatic
evolution $\theta(t) = 0\rightarrow \pi$.\end{itemize}

\textsc{Proof}: Denote the Hamiltonian with a $\Gamma$ flux as
$H_\Gamma(\theta(t))$. Due to the condition $e^{i\pi \Gamma}=-1$,
we know that $H_\Gamma(\pi)$ is the same as the Hamiltonian with a
{\em charge} $\pi$-flux tube. Consequently, two distinct adiabatic
evolutions can be defined between $H_\Gamma(0)$ and
$H_\Gamma(\pi)$, that is, the process $l_\Gamma$ through spin
$\Gamma$ flux threading, and the process $l_c$ through charge flux
threading. The combination of them $l=l_c^{-1}\cdot l_\Gamma$
leads to a closed path in the parameter space. Given the condition
that the system with no flux has a unique ground state, the charge
pumped during such a process must be an integer, denoted as
$N_\Gamma$. Moreover, the charge pumped during the path $l_c^{-1}$
has to be zero, since the Hall conductivity of the system vanishes
due to time-reversal symmetry. Consequently, an integer number of
charge $N_\Gamma e$ is pumped towards the flux tube during the
first half of the adiabatic process, $l_\Gamma$. (Such a
combination of spin and charge flux threading has been introduced
before in Ref.\cite{essin2007}.) Note that the time reversal
symmetry is essential for obtaining the integer charge. In the
integer QH systems, a $\pi$ flux generally induces a fractional
charge of $e/2$ near the flux tube\cite{weeks2007,lee2007}.

\begin{itemize}\item\textsc{Theorem II}: A topological index, defined by
$(-1)^{N_\Gamma}$, is independent of the choice of $\Gamma$, as long
as $e^{i\pi\Gamma}=-1$ is satisfied.\end{itemize}

\textsc{Proof}: For two different operators $\Gamma_1,~\Gamma_2$
satisfying $e^{i\pi \Gamma_{1,2}}=-1$, two different adiabatic
pathes $l_1,l_2$ connecting $\theta=0$ and $\theta=\pi$
Hamiltonians are defined. Consequently, a closed path can be
formed as $l=l_2^{-1}\cdot l_1$, which brings a system without
flux back onto itself. If the number of charge pumped during $l_1$
and $l_2$ is $N_{\Gamma_1}$ and $N_{\Gamma_2}$, respectively, the
net charge pumped during such a process is given by $N_{\rm
tot}=N_{\Gamma_1}-N_{\Gamma_2}$. Since the spin $\Gamma$ flux
preserves time-reversal symmetry, if the initial state
$\left|G\right\rangle$ is a Kramers singlet, so is the final state
$\left|F\right\rangle$. This is only possible if the charge
$N_{\rm tot}$ pumped during the closed path $l=l_2^{-1}\cdot l_1$
is an even integer, leading to the conclusion that
$(-1)^{N_{\Gamma 1}}=(-1)^{N_{\Gamma 2}}$ for any two choices
$\Gamma_1$ and $\Gamma_2$.

Based these two general theorems, we see that the $Z_2$ topological
index $(-1)^{N_\Gamma}$ is independent of the choice of $\Gamma$,
which is thus a well-defined property of the TRI insulator. In this
way, we obtain the following general topological classification for
TRI insulators in 2D.

\begin{itemize}\item\textsc{Definition II}: A topologically trivial TRI insulator is
defined by $(-1)^{N_\Gamma}=1$, and a topologically non-trivial
TRI insulator is defined by $(-1)^{N_\Gamma}=-1$. A topologically
non-trivial TRI insulator displays the quantized spin Hall effect
in the sense that a spin $\Gamma$ flux of $\pi$ pumps an odd
number of quantized electric charges.\end{itemize}

Since the charge pumped during such an adiabatic process is always
well-defined without relying on the detail of the system, such a
general $Z_2$ topological classification is applicable to any
generic many-body system with interaction and disorder. For a
nontrivial insulator with $N_\Gamma$ odd, the adiabatic evolution
$H_{\Gamma}(\theta),~\theta(t)=0\rightarrow \pi$ brings the ground
state $\left|\psi_0\right\rangle$ to a state
$\left|\psi_c\right\rangle$ which is a Kramers singlet but carries
an odd number of electric charge $N_\Gamma e$ around the flux
tube. According to the definition I, $\left|\psi_c\right\rangle$
is a chargeon or holon state. Once the flux is fixed to be $\pi$,
no local perturbation can change the spin-charge separation nature
of the system. In general a local operator $\hat{O}_+$ can be
defined, which acts only around the flux tube and carries charge
$-N_\Gamma e$. When $N_\Gamma$ is odd, $\hat{O}_+$ has to form a
doublet representation of time-reversal transformation ${\cal T}$
together with its partner $\hat{O}_-={\cal T}^{-1}\hat{O}_+{\cal
T}$. Thus the quantum states
$\left|\psi_s^\pm\right\rangle=\hat{O}_\pm\left|\psi_c\right\rangle$
carry vanishing electric charge but form a doublet representation
of ${\cal T}$. According to the definition I, such a pair of state
$\left|\psi_s^\pm\right\rangle$ is a spinon state.

%Mid gap picture for the HgTe model
\begin{figure}[htpb]
    \begin{center}
        \includegraphics[width=3in]{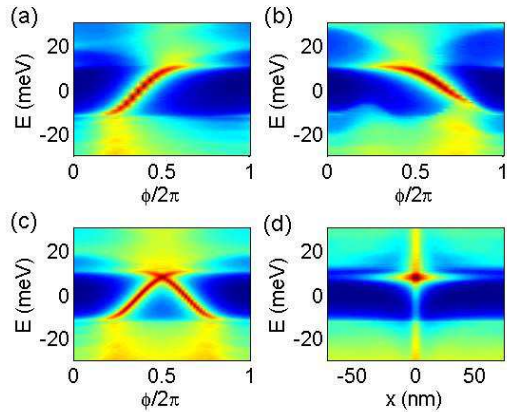}
    \end{center}
    \caption{Local DOS in the core of a flux tube. The dark blue color shows the bulk energy gap and
    the red line shows the mid-gap states. (a) and (b) show the evolution of the mid-gap state upon spin
    $\Gamma$-flux threadings, with $\Gamma=\sigma_z\otimes \mathbb{I}$ and
    $\Gamma=-\left(\sigma_z+\sigma_x\right)\otimes\mathbb{I}/\sqrt{2}$ respectively in the representation used by Ref.\cite{bernevig2006}.
    (c) shows the case of charge flux
    threading. The two mid-gap states cross at $\phi=\pi$. The
    spatial distribution of the mid-gap state at $\phi=\pi$
    is shown in Fig. (d), in which
    the zero of x-axis corresponds to the position of the flux tube. Here and below, all the calculations are done for the
    HgTe model of $d=70\AA$ quantum well, with a lattice constant of
    $a=30\AA$.
    }
    \label{fig:midgapstate}
\end{figure}
To see more explicitly the realization of spinon and holon states
through the adiabatic processes, we can study a typical model of the
topologically nontrivial insulator---the four-band effective
tight-binding model of HgTe quantum well proposed in Ref.
\onlinecite{bernevig2006}. To study the generic case with no $U(1)$
spin symmetry, we have included the bulk-inversion asymmetry (BIA)
terms in our calculation\cite{dai2007}. Both the spin $\Gamma$ and
the charge flux on a plaquette induce mid-gap states inside the gap.
In Fig. \ref{fig:midgapstate}, we show the local density of states
$\rho_i(E)=\sum_n\left|\left\langle
i|n\right\rangle\right|^2\delta\left(E_n-E\right)$ in the core site
$i$ of both the charge and the spin $\Gamma$-flux tubes. As shown in
Fig. \ref{fig:midgapstate}(a) and (b), when a spin $\Gamma$-flux is
threaded from $0$ to $2\pi$, the time-reversal symmetry is preserved
and a Kramers pair of mid-gap states move from the valence band to
the conduction band, or vise versa, depending on the choice of
$\Gamma$ matrix. Consequently, $\pm 2 e$ units of charge are pumped
during the periodic process $\phi=0\rightarrow 2\pi$. The Kramers
double degeneracy of the midgap states demonstrates the key
statement of {\em Theorem II}, {\em i.e.}, that even number of
electric charges are pumped during any time-reversal invariant
cycle. Consequently, from the proof of {\em Theorem II}, we know
that the state at $\phi=\pi$ is a chargeon (holon) for the case of
Fig. \ref{fig:midgapstate} (a) ((b)), respectively. On the other
hand, the spinon states can be obtained by threading a charge $\pi$
flux. As shown in Fig. \ref{fig:midgapstate} (c), a pair of mid-gap
states appear around the charge flux, with a level crossing
occurring at $\phi=\pi$\cite{fu2006b}, as required by time-reversal
symmetry. Consequently, only one of the mid-gap states are occupied
in the final state of such an adiabatic evolution, leading to a
doublet of spinon states.

\begin{figure}[htpb]
    \begin{center}
        \includegraphics[width=1.5in]{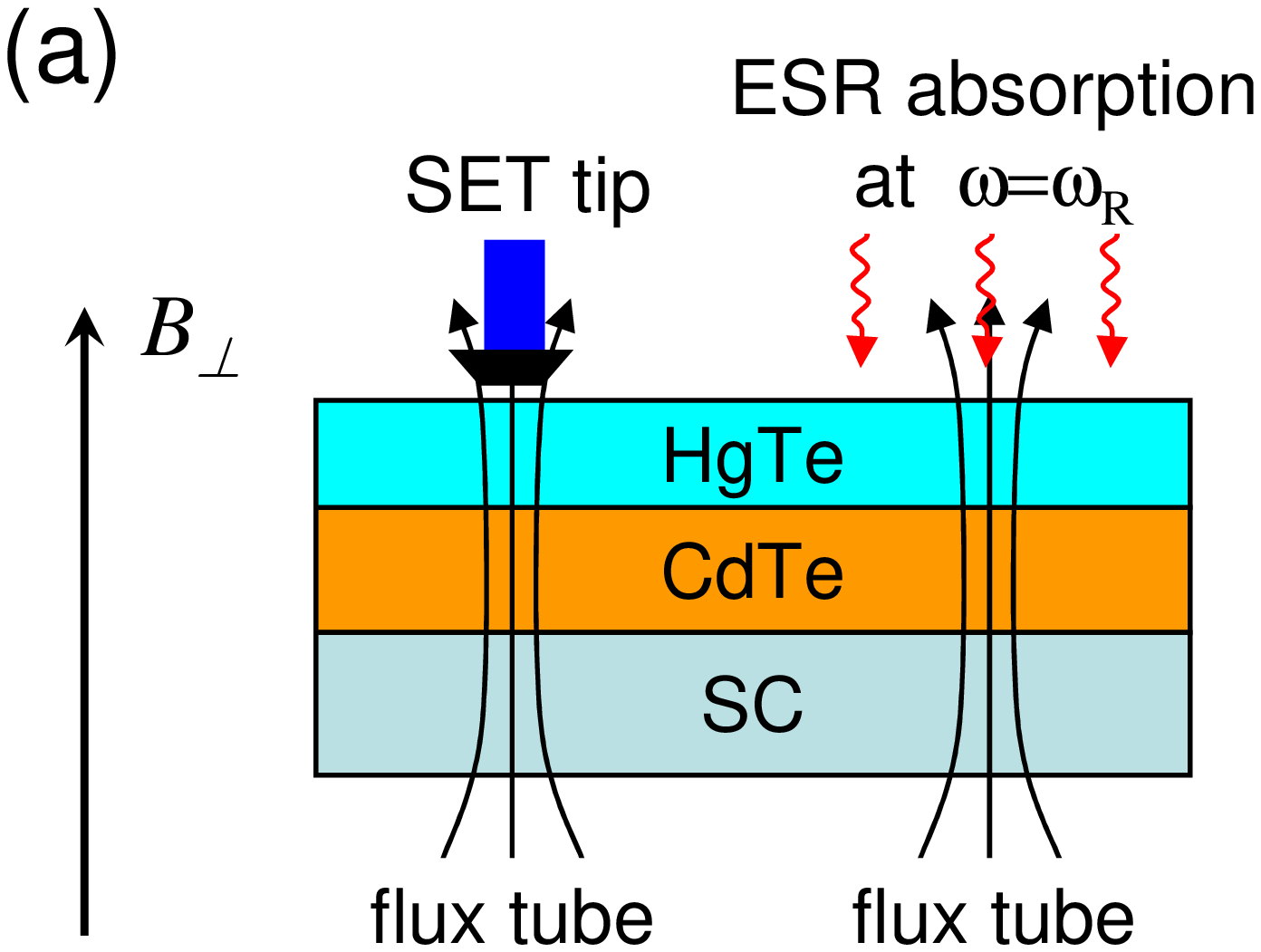}\includegraphics[width=1.5in]{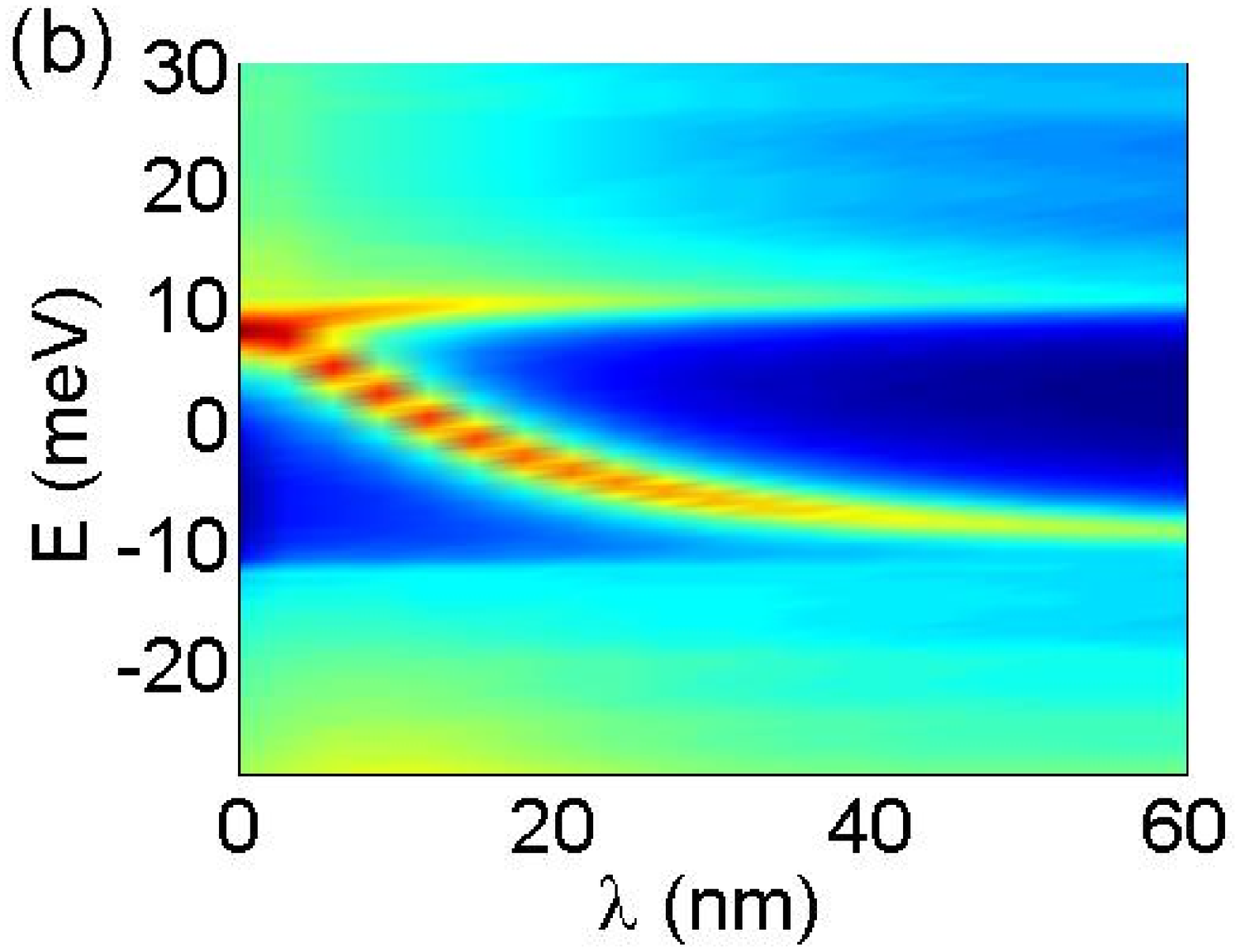}\\
        \includegraphics[width=1.5in]{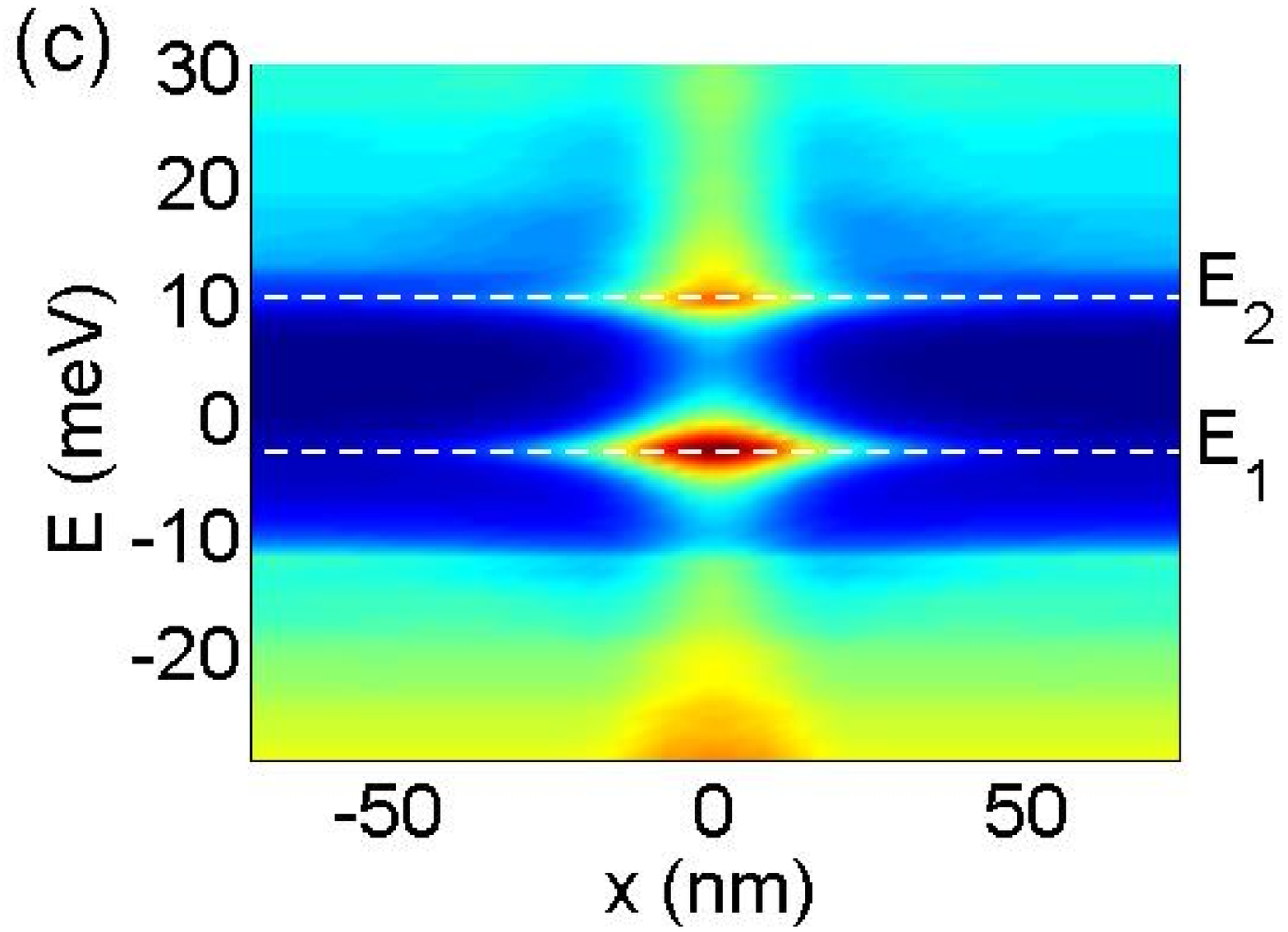}\includegraphics[width=1.5in]{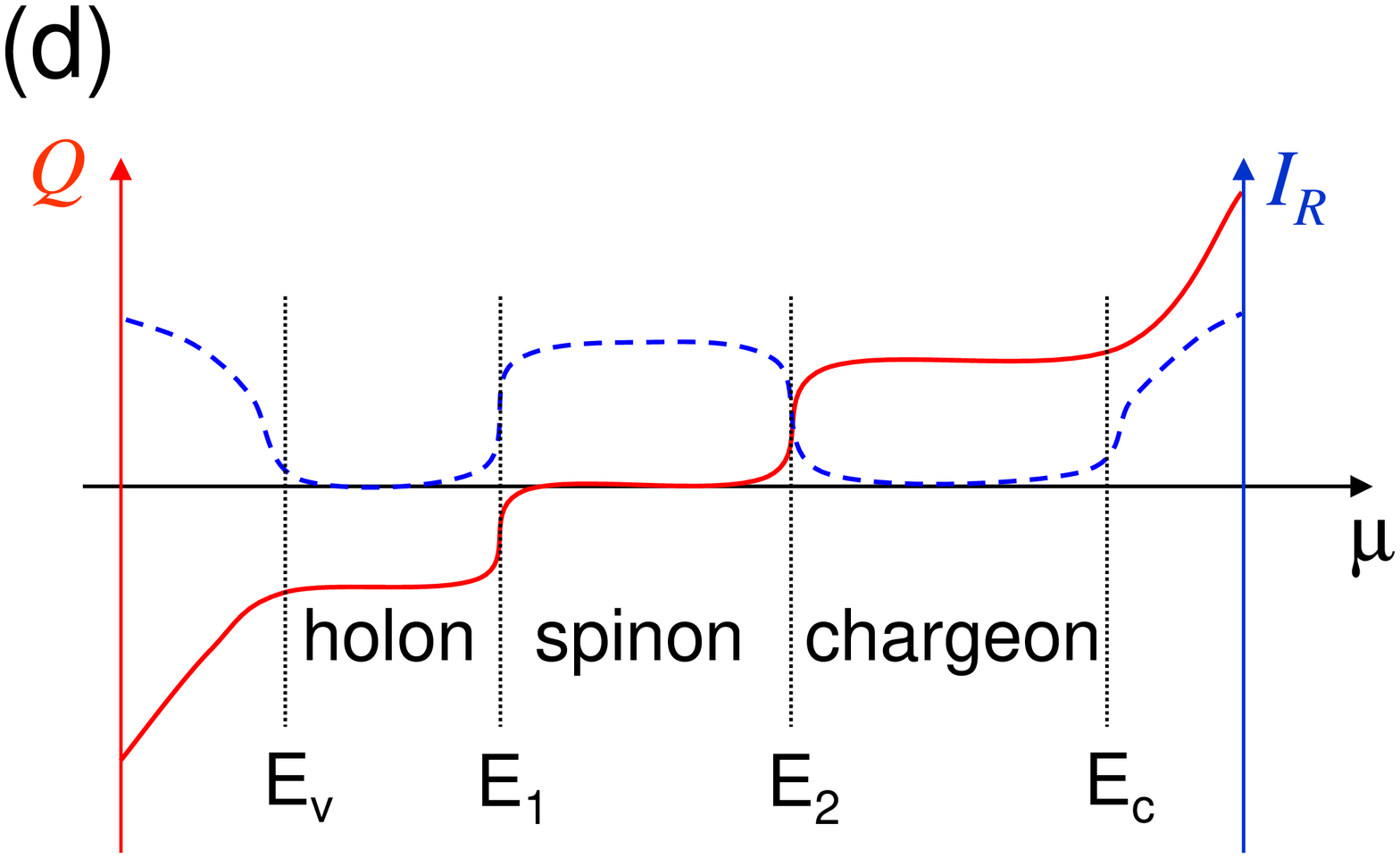}
    \end{center}
    \caption{(a) The superconductor-quantum
    well hybrid structure, with the $\pi$ flux tubes generated by a perpendicular magnetic
    field. (b) The splitting of two mid-gap states upon increasing
    $\lambda$.
    (c) The spatial distribution of the two mid-gap
    states for $\lambda=18{\rm nm}$, in which the dash lines mark the energy
    $E_1$ and $E_2$ of the mid-gap
    states. (d) Illustration of the measured
    charge $Q$ (red line) and intensity of ESR signal $I_R=I(\omega_R)$
    at resonance frequency $\omega_R=(E_2-E_1)/\hbar$ (blue dash line) as
    functions of the chemical potential.
    }
    \label{fig:experiment}
\end{figure}

We now discuss the experimental realization of spin-charge
separation. Consider a hybrid structure between a type-II
superconductor and the HgTe quantum well as shown in Fig.
\ref{fig:experiment}(a). $hc/2e$ or $\pi$ flux tubes are created
by a perpendicular magnetic field $H_{c1}<H<H_{c2}$, with
$H_{c1,2}$ being the lower and upper critical fields of the
superconductor. The superconducting flux tube has a finite size
determined by the penetration depth $\lambda$. In this case, the
time-reversal symmetry is broken even if the net flux is $\pi$. As
shown in Fig. \ref{fig:experiment}(b), the two mid-gap states are
split with increasing $\lambda$. However, at a Gaussian loop with
radius $r_G\gg \lambda$, there is no observable difference between
such a realistic $\pi$ flux and an ideal $\pi$ flux threading into
a plaquette, and the spinon and holon/chargeon states still exist
and remain topologically distinct from a trivial many-body state
of the electron system. Denoting $E_1$ and $E_2$ the energy of the
two mid-gap states, and $E_v$ and $E_c$ the energy of valence band
top and conduction band bottom, at zero temperature the ground
state of the system is given by: (i) the holon state, when the
chemical potential $E_v<\mu<E_1$; (ii) the spinon state with a
preferred spin in the magnetic field, when $E_1<\mu<E_2$; (iii)
the chargon state, when $E_2<\mu<E_c$. Consequently, the
spin-charge separation can be observed if we can measure the
charge and spin induced by a flux tube independently. The local
charge distribution in the chargeon or holon state can be probed
by the recently developed scanning single-electron transistor
(SET)\cite{yoo1997}, while the spin carried by the spinon state
can be observed by electron spin resonance (ESR). Only when the
system is in the spinon state, a transition between the two
mid-gap states can occur, leading to a sharp resonance peak in the
ESR signal at frequency $\omega_R=(E_2-E_1)/\hbar$. The
qualitative behavior of local charge and ESR spectrum is
summarized in Fig.\ref{fig:experiment} (d), as a function of the
chemical potential, which can be tuned by the back gate voltage.
Moreover, it can be seen from Fig.\ref{fig:experiment} (b) that
the energy splitting $E_2-E_1$ is proportional to $\lambda$ for
small $\lambda$. Since the penetration depth of a superconductor
film depends on the thickness of the film, the resonance frequency
$\omega_R$ can be measured for several different $\lambda$, which
should extrapolate to $\omega_R\rightarrow 0$ in the limit
$\lambda\rightarrow0$. If observed, such an asymptotic behavior
can demonstrate the existence of a Kramers pair in the ideal case
with time-reversal symmetry.

In conclusion, we have given a $Z_2$ classification of TRI
insulators which is generally valid in the presence of interactions
and disorder. We showed that this topological property can be
measured experimentally by the spin-charge separation in the
presence of a $\pi$ flux. We provided an experimental proposal to
observe the spin-charge separation in an SC-QSH hybrid structure.
Fractionalization is usually accompanied by fractional statistics.
By studying the topological effective theory of the SC-QSH hybrid
system, it can be shown that the spinon, holon and chargeon are all
bosons, but each spinon acts as a $\pi$-flux for chargeon and holon,
and vise versa. In other words, this system has nontrivial {\em
mutual statistics} which is described by a mutual Chern-Simons
theory.\cite{kou2005}

Such a relation between spin-charge separation and TRI topological
insulators can also be generalized to higher dimensions. In a 3D TRI
insulator with a $\pi$-flux ring, {\em Theorem I} and {\em II} can
be generalized straightforwardly, which results in a generic $Z_2$
classification of 3D TRI insulators. In a three-d nontrivial
insulator (which corresponds to the ``strong topological insulator"
defined in literature\cite{moore2007a,fu2007c,roy2006a}), a closed
string with $\pi$ flux is a spin-charge separated extended object,
satisfying our {\em Definition I} of spin-charge separation.

During the course of this work, we became aware of the independent
work by Y. Ran, A. Vishwanath and D.-H. Lee, demonstrating
spin-charge separation in similar models, which is now posted in
arXiv:0801.0627 \cite{ran2008}. We acknowledge helpful discussions
with T. Hughes, C. L. Kane, S. A. Kivelson, L. Molenkamp and Cenke
Xu. This work is supported in part by the NSF through the grants
DMR-0342832, and by the US Department of Energy, Office of Basic
Energy Sciences under contract DE-AC03-76SF00515.

\bibliography{SCS,QSH}
\end{document}